# Generic bond energy formalism within the modified quasichemical model for ternary solutions

Kun Wang[1*], Dongyang Li[2], Xingli Zou[1], Hongwei Cheng[1], Chonghe Li[1], Xionggang Lu[1], Kuochih Chou[1]

1 State Key Laboratory of Advanced Special Steel & Shanghai Key Laboratory of Advanced Ferrometallurgy & School of Materials Science and Engineering, Shanghai University, 333 Nanchen Road, Baoshan District, Shanghai 200444, China

2 Department of Chemical and Materials Engineering, University of Alberta, Edmonton, Canada T6G 1H9

**Abstract**: The Modified Quasichemical Model in the Pair Approximation (MQMPA) can effectively capture the thermodynamic features of a binary solution with Short-Range Ordering (SRO). If the model is used to treat a ternary solution, a geometric interpolation method must be employed to extend the bond energy expression from binary to ternary formalism. The aim of the present work is to implement such extension by means of a generic geometric interpolation approach. The generic method is unbiased and can be transformed into the widely used Kohler, Toop and Muggianu approaches with special interpolation parameters. The interpolation parameters can be calculated by the integration method as well as be optimized by ternary experimental data. The generic bond energy formalism (GBEF) has thus been derived to provide the MQMPA great flexibility to describe ternary solutions with complex configurations. Moreover, the GBEF is more concise than the formula derived by a combinatorial Kohler-Toop method. The concise GBEF is in the respect more conveniently programmed. Eventually, the Cu-Li-Sn liquid where both SRO and clustering among atoms occur is employed to validate the effectiveness and reliability of the GBEF within the MQMPA.

## 1. Introduction

Thermodynamics and phase diagrams have provided strict laws and methods [1] to understand and investigate the interrelationship among compositions, properties, phase equilibria, microstructures and process conditions. They are thus metaphorized as the "GPS" navigation

[*]Corresponding author

E-mail address: wangkun0808@shu.edu.cn

system for material exploration, design and preparation. Nowadays, the research paradigm for phase diagrams has greatly shifted from pure experimental measurements to theoretical predictions incorporated with key experimental validations, especially for multicomponent systems. Owing to this transformation, the investigations of multicomponent phase diagrams are thus booming because the new paradigm can effectively reduce the experimental measurements that are expensive and time-consuming.

The currently prevailing theoretical framework for predicting phase diagrams is the so-called Calphad (Calculation of phase diagrams) method [2-5]. This method employs the thermodynamic information from binary and ternary solutions to predict that of multicomponent solutions and then calculate multicomponent phase diagrams. In the near future, the Calphad method is still the sole way to efficiently compute multicomponent multi-phase equilibria since thermodynamic information is experimentally or theoretically available for most low-order systems while atomistic simulations are rather expensive in terms of computational resources. The accuracy of the Calphad method depends largely upon the reliability of a thermodynamic model for multicomponent solutions. The reliability involves two aspects. Firstly, the thermodynamic model must be reliable enough for characterizing the Gibbs energy of each constituent binary solution no matter whether the solution configuration presents short-range ordering (SRO). Secondly, the thermodynamic model must have a generic geometrical interpolation approach that is unbiased, changeable and system-dependable to predict the thermodynamic properties of ternary or higher-order solutions from those of all the constituent binary solutions. It seems that the two aspects are not well achieved by the Bragg-Williams-Muggianu (BWM) model [6-7] although it is widely used to treat a vast amount of multicomponent solutions.

The present paper is devoted to the development of a new thermodynamic model for ternary solutions. The new model is built based on the MQMPA and GBEF. The MQMPA [8] is widely used in the Factsage community to treat binary solutions no matter whether they appear SRO. It is thought to be the right response to the first remark of the model reliability. Pelton et al. [9] also proposed a bond energy expression applicable to ternary solutions by means of a combinatorial Kohler-Toop method. However, it requires human interference to arrange components to three apexes in the Kohler-Toop triangle according to some empirical rules, such as chemical properties of components, locations of elements in the periodic table, valences of compounds, etc.

In addition, the Kohler-Toop interpolation renders the bond energy formalism quite complex, especially for multicomponent systems. The general method requires no human interference to make the arrangement. It instead introduces interpolation parameters that can be adjusted using ternary experimental data. If no ternary data is available, these parameters may also be reasonably calculated by the integration method [10], which helps address possible issues related to the second remark of the model reliability regarding how to achieve a generic, unbiased and system-dependable interpolation scheme. These interpolation parameters can be temperature dependent and within the range from 0 to 1. The codomain enables the generic interpolation to be ergodic over all the Kohler [11], Toop [12] and Muggianu [13] approaches. Moreover, a more concise formalism can be derived by the generic method for expressing the bond energy in ternary solutions. This allows convenient model comprehension and code implementation. The present model is able to characterize ternary solutions no matter whether SRO occurs in the constituent binary solutions. Moreover, it requires no empirical selection of the interpolation methods. In this study, the Cu-Li-Sn liquid is employed to validate the present model in section 4, since complex solution configurations occur over the entire composition zone. The complicated configurations result from a strong SRO in the Li-Sn liquid and respective attractive and repulsive interactions between alloying atoms in the Cu-Sn and Li-Cu liquids. In a follow-up study [14], the GBEF would be derived to handle multicomponent solutions within the MQMPA.

## 2. The Model

Within the framework of the MQMPA [9], the Gibbs energy of a ternary solution can be expressed as

$$G = (n_A g_A^0 + n_B g_B^0 + n_C g_C^0) + RT \left( n_A \ln X_A + n_B \ln X_B + n_C \ln X_C + n_{AA} \ln \left( \frac{X_{AA}}{Y_A^2} \right) + n_{BB} \ln \left( \frac{X_{BB}}{Y_B^2} \right) + n_{CC} \ln \left( \frac{X_{CC}}{Y_C^2} \right) + n_{AB} \ln \left( \frac{X_{AB}}{2Y_A Y_B} \right) + n_{AC} \ln \left( \frac{X_{AC}}{2Y_A Y_C} \right) + n_{BC} \ln \left( \frac{X_{BC}}{2Y_B Y_C} \right) \right) + \left( n_{AB} \frac{\Delta g_{AB}}{2} + n_{AC} \frac{\Delta g_{AC}}{2} + n_{BC} \frac{\Delta g_{BC}}{2} \right) \quad (1)$$

where $g_A^0$, $n_A$, $X_A$ and $Y_A$ are respectively the standard molar Gibbs energy, the number of moles, the mole fraction, and the "coordination-equivalent" fraction of component A (similarly for components B and C). $n_{AA}$ and $X_{AA}$ are respectively the number of moles and the mole fraction of pair (A-A) (similarly for pairs (B-B), (C-C), (A-B), (A-C) and (B-C)). In the following expressions, relative definitions will only be given for component A and pair (A-A). Readers should figure out how to similarly define other components and pairs. $R$ and $T$ refer to the gas constant and temperature in Kelvin, respectively. In Equation (1), the first term is the linear summation of the Gibbs energies from the pure components, the second term is the Gibbs energy contributed by the configurational entropy of mixing, and the last term is the total energy required or released to form all the A-B, A-C and B-C pairs in the ternary solution. The pair exchange reaction can be schematically represented as

$$(A-A) + (B-B) = 2(A-B) \quad \Delta g_{AB} \tag{2}$$

where $\Delta g_{AB}$ refers to the total energy required or released to form two moles of the A-B pairs. There are in total three kinds of such pair reactions in the A-B-C solution. All these represent the first-nearest-neighbor (FNN) pairs. In the binary A-B solution, $\Delta g_{AB}$ can be expressed either in terms of the coordination-equivalent fraction as

$$\Delta g_{AB} = \Delta g_{AB}^o + \sum_{i+j \geq 1} g_{AB}^{ij} Y_A^i Y_B^j \tag{3}$$

or in terms of the pair fraction as

$$\Delta g_{AB} = \Delta g_{AB}^o + \sum_{i+j \geq 1} g_{AB}^{ij} X_{AA}^i X_{BB}^j \tag{4}$$

where $\Delta g_{AB}^0$ and $g_{AB}^{ij}$ are the model parameters that can be functions of temperature. $\Delta g_{AB}$ and $\Delta g_{AC}$ have the similar definitions.

The interrelations among those substance quantities in Equation (1) are defined as

$$X_A = \frac{n_A}{n_A + n_B + n_C} \tag{5}$$

$$Z_A n_A = 2 n_{AA} + n_{AB} + n_{AC} \tag{6}$$

$$X_{AA} = \frac{n_{AA}}{n_{AA} + n_{BB} + n_{CC} + n_{AB} + n_{AC} + n_{BC}} \tag{7}$$

$$Y_A = \frac{Z_A n_A}{Z_A n_A + Z_B n_B + Z_C n_C} = \frac{Z_A X_A}{Z_A X_A + Z_B X_B + Z_C X_C} = X_{AA} + \frac{1}{2}(X_{AB} + X_{AC}) \tag{8}$$

where $Z_A$ is the overall coordination number of component A in the A-B-C solution. For an earlier version of the MQMPA, $Z_A$ is a constant value and independent of composition. However, the immutable coordination number brings a number of drawbacks [8]. In the current MQMPA [8], $Z_A$ is allowed to vary with composition:

$$\frac{1}{Z_A} = \frac{1}{2n_{AA} + n_{AB} + n_{AC}} \left( \frac{2n_{AA}}{Z_{AA}^A} + \frac{n_{AB}}{Z_{AB}^A} + \frac{n_{AC}}{Z_{AC}^A} \right) \tag{9}$$

where $Z_{AA}^A$ and $Z_{AB}^A$ are the values of $Z_A$ when all the nearest neighbors of an A are As and when all the nearest neighbors of an A are Bs, respectively. Substituting Equation (9) into Equation (6) gives

$$n_A = \frac{2n_{AA}}{Z_{AA}^A} + \frac{n_{AB}}{Z_{AB}^A} + \frac{n_{AC}}{Z_{AC}^A} \tag{10}$$

Equation (10) is the true mass balance between the number of moles of component A and the numbers of moles of pairs (A-A), (A-B) and (A-C) in the ternary solution if these pairs are formally treated as the fractional associates, $A_{1/Z_{AA}^A}A_{1/Z_{AA}^A}$, $A_{1/Z_{AB}^A}B_{1/Z_{AB}^B}$ and $A_{1/Z_{AC}^A}C_{1/Z_{AC}^C}$. In this regard, Equation (1) can be reformed as

$$G = (n_{AA}g_{AA}^o + n_{BB}g_{BB}^o + n_{CC}g_{CC}^o + n_{AB}g_{AB}^o + n_{AC}g_{AC}^o + n_{BC}g_{BC}^o) + RT(n_{AA}\ln X_{AA} + n_{BB}\ln X_{BB} + n_{CC}\ln X_{CC} + n_{AB}\ln X_{AB} + n_{AC}\ln X_{AC} + n_{BC}\ln X_{BC}) + RT(n_A \ln X_A + n_B \ln X_B + n_C \ln X_C - n_{AA}\ln Y_A^2 - n_{BB}\ln Y_B^2 - n_{CC}\ln Y_C^2 - n_{AB}\ln(2Y_A Y_B) - n_{AC}\ln(2Y_A Y_C) - n_{BC}\ln(2Y_B Y_C)) + \left(n_{AB}\left(\frac{\Delta g_{AB}-\Delta g_{AB}^o}{2}\right) + n_{AC}\left(\frac{\Delta g_{AC}-\Delta g_{AC}^o}{2}\right) + n_{BC}\left(\frac{\Delta g_{BC}-\Delta g_{BC}^o}{2}\right)\right) \tag{11}$$

where $g_{AA}^o$ and $g_{AB}^o$ are the standard molar Gibbs energies of pairs (A-A) and (A-B), respectively. They can be defined according to the mass-balance principle as

$$g_{AA}^o = \frac{2}{Z_{AA}^A} g_A^o \tag{12}$$

$$g_{AB}^o = \frac{1}{Z_{AB}^A} g_A^o + \frac{1}{Z_{AB}^B} g_B^o + \frac{\Delta g_{AB}^o}{2} \tag{13}$$

where $\Delta g_{AB}^o$ is the binary parameter from Equations (3) or (4). In comparison with the Associate Solution Model (ASM), the quasichemical model has an extra entropy term, which is the third part in Equation (11). It is this part that resolves the "entropy paradox" [15] of the ASM and allows the MQMPA to reduce to an ideal solution model in the limit. Moreover, the same existing algorithms and computer subroutines commonly used for the ASM can be conveniently applied to the MQMPA with minor modifications.

The model parameters in Equations (3) and (4) are optimized from binary experimental data in thermochemistry and phase equilibria. Equations (3) and (4) can only apply to the sub-binary solution. It is now required to derive the bond energy ($\Delta g_{AB}$) expressions in terms of either the "coordination-equivalent" fractions or the pair fractions for dealing with ternary solutions. Similar treatments are also applied to $\Delta g_{AC}$ and $\Delta g_{BC}$. The detailed derivations are given in the next section.

## 3. The interpolation formulae

At the beginning of this section, the binary bond energy expressions ($\Delta g_{AB}$) written in the polynomials of the "coordination-equivalent" fractions and the pair fractions are both expanded for dealing with the ternary A-B-C solution via a generic interpolation method. The resultant GBEFs are then explored for some special conditions under which they are equivalent to those expanded by the Kohler, Toop and Muggianu methods.

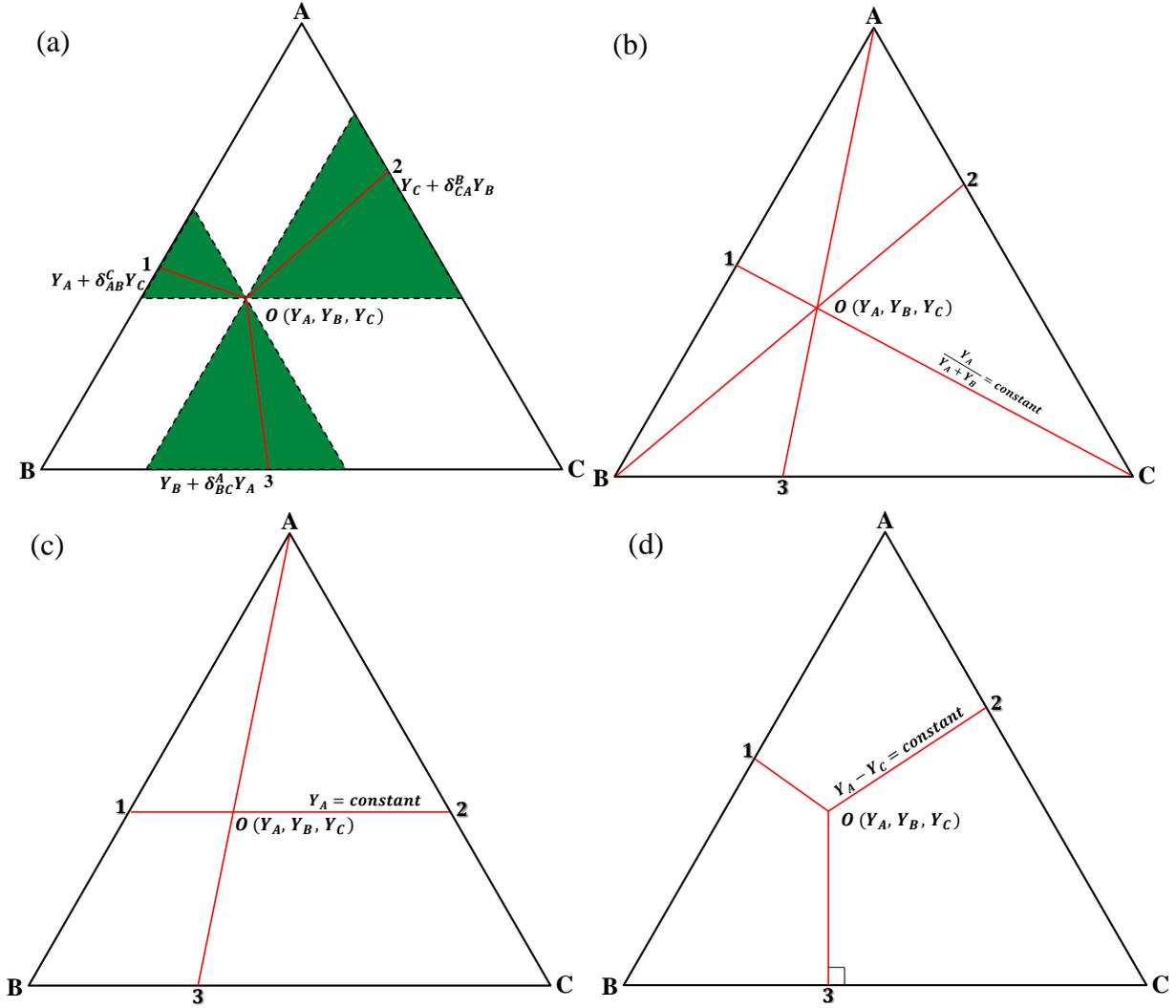

Fig. 1. Some geometrical solution models: (a) generic interpolation; (b) Kohler interpolation; (c) Kohler-Toop interpolation; (d) Muggianu interpolation

## 3.1 The generic method

Assuming that $\Delta g_{AB}$ is written in Equation (3), the generic interpolation method, as illustrated in Fig.1, can transform the bond energy formalism for the binary A-B solution to that for the ternary A-B-C solution as

$$\Delta g_{AB} = \left(\Delta g_{AB}^0 + \sum_{(i+j)\geq 1} q_{AB}^{ij} (Y_A + \delta_{AB}^C Y_C)^i (Y_B + \delta_{BA}^C Y_C)^j\right) + \sum_{\substack{k\geq 1 \\ i\geq 0 \\ j\geq 0}} q_{AB(C)}^{ijk} (Y_A + \delta_{AB}^C Y_C)^i (Y_B + \delta_{BA}^C Y_C)^j Y_C^k \quad (14)$$

where the first term on the right-hand side is constant along the line "1-$O$" in Fig.1a and is equal to $\Delta g_{AB}$ at the A-B side at point 1 (where $\delta_{AB}^C + \delta_{BA}^C = 1$). This means the (A-B) bond energy keeps constant along the "1-$O$" line. The position of "1" is chosen according to the interpolation parameter $\delta_{AB}^C$, which can be a function of temperature, and whose value is within the range from 0 to 1. The codomain of $\delta_{AB}^C$ enables the generic approach to be ergodic over all the reported geometrical interpolation methods. The value of $\delta_{AB}^C$ can be calculated using the integration method [10]. It can also be optimized by ternary experimental data. Similar treatments can be used for $\delta_{BC}^A$ and $\delta_{CA}^B$. The second summation in Equation (14) contains ternary terms that are all zero in the A-B binary system, which give the effect of the presence of component C upon the (A-B) bond energy. The empirical ternary coefficients $q_{AB(C)}^{ijk}$ can also be optimized by ternary experimental data. Actually, ternary coefficients could be very small if interpolation parameters are appropriate enough.

The generic ternary $\Delta g_{CA}$ and $\Delta g_{BC}$ expressions are similarly given as

$$\Delta g_{CA} = \left(\Delta g_{CA}^0 + \sum_{(i+j)\geq 1} q_{CA}^{ij}(Y_C + \delta_{CA}^B Y_B)^i (Y_A + \delta_{AC}^B Y_B)^j\right) + \sum_{\substack{k\geq 1 \\ i\geq 0 \\ j\geq 0}} q_{CA(B)}^{ijk}(Y_C + \delta_{CA}^B Y_B)^i (Y_A + \delta_{AC}^B Y_B)^j Y_B^k \quad (15)$$

$$\Delta g_{BC} = \left(\Delta g_{BC}^0 + \sum_{(i+j)\geq 1} q_{BC}^{ij}(Y_B + \delta_{BC}^A Y_A)^i (Y_C + \delta_{CB}^A Y_A)^j\right) + \sum_{\substack{k\geq 1 \\ i\geq 0 \\ j\geq 0}} q_{BC(A)}^{ijk}(Y_B + \delta_{BC}^A Y_A)^i (Y_C + \delta_{CB}^A Y_A)^j Y_A^k \quad (16)$$

where the (C-A) and (B-C) bond energies (the first terms at the right-hand sides) are equal to their values at point "2" (where $\delta_{CA}^B + \delta_{AC}^B = 1$) and "3" (where $\delta_{BC}^A + \delta_{CB}^A = 1$), respectively, as shown in Fig.1a. The ternary coefficients $q_{CA(B)}^{ijk}$ and $q_{BC(A)}^{ijk}$ in $\Delta g_{CA}$ and $\Delta g_{BC}$ give the effects of component B and A upon the (C-A) and (B-C) bond energies, respectively. It is considered [16] that the functional form for the ternary terms is preferable, compared with traditional expressions such as $Y_A^i Y_B^j Y_C^k$.

Suppose $\Delta g_{AB}$ is expressed as Equation (4). Correspondingly, the GBEF for the ternary solution can be formulated as

$$\Delta g_{AB} = \Delta g_{AB}^0 + \sum_{(i+j)\geq 1} q_{AB}^{ij}(X_{AA} + {\delta_{AB}^C}^2 X_{CC} + \delta_{AB}^C X_{AC})^i (X_{BB} + {\delta_{BA}^C}^2 X_{CC} + \delta_{BA}^C X_{BC})^j$$

$$+\sum_{\substack{k\geq 1\\i\geq 0\\j\geq 0}} q^{ijk}_{AB(C)} (X_{AA} + \delta^{C}_{AB}{}^2 X_{CC} + \delta^{C}_{AB} X_{AC})^i (X_{BB} + \delta^{C}_{BA}{}^2 X_{CC} + \delta^{C}_{BA} X_{BC})^j Y^k_C \qquad (17)$$

where $\Delta g^0_{AB}$ and $q^{ij}_{AB}$ are optimized by binary experimental data, and $q^{ijk}_{AB(C)}$ is determined by ternary experimental data. The formalism is derived by considering the following facts. As $\Delta g_{AB}$, $\Delta g_{BC}$ and $\Delta g_{CA}$ become small, the solution approaches ideality, leading to $X_{AA} \to Y^2_A$, $X_{BB} \to Y^2_B$, $X_{CC} \to Y^2_C$, $X_{AB} \to 2Y_A Y_B$, $X_{AC} \to 2Y_A Y_C$ and $X_{BC} \to 2Y_B Y_C$. In this case, Equation (17) approaches Equation (14) since $X_{AA} + \delta^{C}_{AB}{}^2 X_{CC} + \delta^{C}_{AB} X_{AC} \to (Y_A + \delta^{C}_{AB} Y_C)^2$ and $X_{BB} + \delta^{C}_{BA}{}^2 X_{CC} + \delta^{C}_{BA} X_{AC} \to (Y_B + \delta^{C}_{BA} Y_C)^2$ are realized.

Following the similar route, the GBEFs of $\Delta g_{CA}$ and $\Delta g_{BC}$ for the ternary solution can be derived as

$$\Delta g_{CA} = \Delta g^0_{CA} + \sum_{(i+j)\geq 1} q^{ij}_{CA} (X_{CC} + \delta^{B}_{CA}{}^2 X_{BB} + \delta^{B}_{CA} X_{BC})^i (X_{AA} + \delta^{B}_{AC}{}^2 X_{BB} + \delta^{B}_{AC} X_{AB})^j$$

$$+\sum_{\substack{k\geq 1\\i\geq 0\\j\geq 0}} q^{ijk}_{CA(B)} (X_{CC} + \delta^{B}_{CA}{}^2 X_{BB} + \delta^{B}_{CA} X_{BC})^i (X_{AA} + \delta^{B}_{AC}{}^2 X_{BB} + \delta^{B}_{AC} X_{AB})^j Y^k_B \qquad (18)$$

$$\Delta g_{BC} = \Delta g^0_{BC} + \sum_{(i+j)\geq 1} q^{ij}_{BC} (X_{BB} + \delta^{A}_{BC}{}^2 X_{AA} + \delta^{A}_{BC} X_{AB})^i (X_{CC} + \delta^{A}_{CB}{}^2 X_{AA} + \delta^{A}_{CB} X_{AC})^j$$

$$+\sum_{\substack{k\geq 1\\i\geq 0\\j\geq 0}} q^{ijk}_{BC(A)} (X_{BB} + \delta^{A}_{BC}{}^2 X_{AA} + \delta^{A}_{BC} X_{AB})^i (X_{CC} + \delta^{A}_{CB}{}^2 X_{AA} + \delta^{A}_{CB} X_{AC})^j Y^k_A \qquad (19)$$

where Equations (18) and (19) approach Equations (15) and (16), respectively, since $X_{CC} + \delta^{B}_{CA}{}^2 X_{BB} + \delta^{B}_{CA} X_{BC} \to (Y_C + \delta^{B}_{CA} Y_B)^2$ along with $X_{AA} + \delta^{B}_{AC}{}^2 X_{BB} + \delta^{B}_{AC} X_{AB} \to (Y_A + \delta^{B}_{AC} Y_B)^2$, and $X_{BB} + \delta^{A}_{BC}{}^2 X_{AA} + \delta^{A}_{BC} X_{AB} \to (Y_B + \delta^{A}_{BC} Y_A)^2$ together with $X_{CC} + \delta^{A}_{CB}{}^2 X_{AA} + \delta^{A}_{CB} X_{AC} \to (Y_C + \delta^{A}_{CB} Y_A)^2$ are envisaged.

From Equations (14) and (17), it can be seen [9] that the ternary terms contain the factor $Y_C$ which is equal to $X_{CC} + (X_{CA} + X_{BC})/2$ from Equation (8). In principle, the effects of the three pair fractions on $\Delta g_{AB}$ could be easily represented by three independent ternary coefficients. However, it may be unnecessary to have this additional complexity. The similar reason is for keeping the factor $Y_A$ in Equations (16, 19) and $Y_B$ in Equations (15, 18).

## 3.2 The Kohler method

As stated in aforementioned sections, the generic interpolation method can be transformed into all the reported approaches when the interpolation parameter is assigned to some special values. When $\delta_{AB}^C$ is defined as $Y_A/(Y_A + Y_B)$, the generic method reduces to the Kohler approach. Substituting it into Equation (14) gives

$$\Delta g_{AB} = \left(\Delta g_{AB}^0 + \sum_{i+j\geq 1} q_{AB}^{ij} \left(\frac{Y_A}{Y_A+Y_B}\right)^i \left(\frac{Y_B}{Y_A+Y_B}\right)^j\right) + \sum_{\substack{k\geq 1 \\ i\geq 0 \\ j\geq 0}} q_{AB(C)}^{ijk} \left(\frac{Y_A}{Y_A+Y_B}\right)^i \left(\frac{Y_B}{Y_A+Y_B}\right)^j Y_C^k \quad (20)$$

where the first term at the right-hand side is constant along the line "1-C" in Fig.1b and is equal to $\Delta g_{AB}$ in the binary system at point "1". This means that the (A-B) bond energy is constant at a fixed $Y_A/Y_B$ ratio. The second term represents the ternary interaction that is well explained previously.

If the Kohler method is also used to derive the ternary $\Delta g_{CA}$ and $\Delta g_{BC}$ expressions, $\delta_{CA}^B$ and $\delta_{BC}^A$ in Equations (15-16) should be replaced by $Y_C/(Y_C + Y_A)$ and $Y_B/(Y_B + Y_C)$, respectively. The substitutions yield

$$\Delta g_{CA} = \left(\Delta g_{CA}^0 + \sum_{i+j\geq 1} q_{CA}^{ij} \left(\frac{Y_C}{Y_C+Y_A}\right)^i \left(\frac{Y_A}{Y_C+Y_A}\right)^j\right) + \sum_{\substack{k\geq 1 \\ i\geq 0 \\ j\geq 0}} q_{CA(B)}^{ijk} \left(\frac{Y_C}{Y_C+Y_A}\right)^i \left(\frac{Y_A}{Y_C+Y_A}\right)^j Y_B^k \quad (21)$$

$$\Delta g_{BC} = \left(\Delta g_{BC}^0 + \sum_{i+j\geq 1} q_{BC}^{ij} \left(\frac{Y_B}{Y_B+Y_C}\right)^i \left(\frac{Y_C}{Y_B+Y_C}\right)^j\right) + \sum_{\substack{k\geq 1 \\ i\geq 0 \\ j\geq 0}} q_{BC(A)}^{ijk} \left(\frac{Y_B}{Y_B+Y_C}\right)^i \left(\frac{Y_C}{Y_B+Y_C}\right)^j Y_A^k \quad (22)$$

where the (C-A) and (B-C) bond energies are constant along the lines "2-B" and "3-C" with constant $Y_C/(Y_C + Y_A)$ and $Y_B/(Y_B + Y_C)$ ratios, and are equal to their values at point "2" and "3" in Fig.1(b), respectively. The second terms in Equations (21-22) represent ternary interactions as before.

Suppose that binary $\Delta g_{AB}$, $\Delta g_{CA}$ and $\Delta g_{BC}$ expressions have been written in Equation (4). With considering the facts as described below Equation (17), their ternary formalisms are expressed by the following equations,

$$\Delta g_{AB} = \left(\Delta g_{AB}^0 + \sum_{i+j\geq 1} q_{AB}^{ij} \left(\frac{X_{AA}}{X_{AA}+X_{BB}+X_{AB}}\right)^i \left(\frac{X_{BB}}{X_{AA}+X_{BB}+X_{AB}}\right)^j\right)$$
$$+ \sum_{\substack{k\geq 1 \\ i\geq 0 \\ j\geq 0}} q_{AB(C)}^{ijk} \left(\frac{X_{AA}}{X_{AA}+X_{BB}+X_{AB}}\right)^i \left(\frac{X_{BB}}{X_{AA}+X_{BB}+X_{AB}}\right)^j Y_C^k \quad (23)$$

$$\Delta g_{CA} = \left(\Delta g_{CA}^0 + \sum_{i+j\geq 1} q_{CA}^{ij} \left(\frac{X_{CC}}{X_{CC}+X_{AA}+X_{CA}}\right)^i \left(\frac{X_{AA}}{X_{CC}+X_{AA}+X_{CA}}\right)^j\right)$$
$$+ \sum_{\substack{k\geq 1 \\ i\geq 0 \\ j\geq 0}} q_{CA(B)}^{ijk} \left(\frac{X_{CC}}{X_{CC}+X_{AA}+X_{CA}}\right)^i \left(\frac{X_{AA}}{X_{CC}+X_{AA}+X_{CA}}\right)^j Y_B^k \quad (24)$$

$$\Delta g_{BC} = \left( \Delta g_{BC}^0 + \sum_{i+j\geq 1} q_{BC}^{ij} \left( \frac{X_{BB}}{X_{BB} + X_{CC} + X_{BC}} \right)^i \left( \frac{X_{CC}}{X_{BB} + X_{CC} + X_{BC}} \right)^j \right)$$
$$+ \sum_{\substack{k\geq 1 \\ i\geq 0 \\ j\geq 0}} q_{BC(A)}^{ijk} \left( \frac{X_{BB}}{X_{BB}+X_{CC}+X_{BC}} \right)^i \left( \frac{X_{CC}}{X_{BB}+X_{CC}+X_{BC}} \right)^j Y_A^k \qquad (25)$$

Since the three components are treated in the same way, this interpolation method is called as the symmetric Kohler model.

However, there are some systems where one component is chemically different from the other two (*e.g.*, BeF$_2$-LiF-NaF, SiO$_2$-CaO-MgO, S-Fe-Cu, *etc.*). In this situation, it is more appropriate to use the "asymmetric" model (Fig.1c) wherein A is selected as the asymmetric component.

### 3.3 The Kohler-Toop method

For the A-B-C solution, if A is very different from B and C in chemistry, the Kohler-Toop method is more appropriate for the interpolations. When $\delta_{AB}^C$ is assigned zero, the generic method can reduce to the Toop-type model. Substituting it into Equation (14) gives the ternary $\Delta g_{AB}$ expression as

$$\Delta g_{AB} = \left( \Delta g_{AB}^0 + \sum_{i+j\geq 1} q_{AB}^{ij} Y_A^i (Y_B + Y_C)^j \right) + \sum_{\substack{k\geq 1 \\ i\geq 0 \\ j\geq 0}} q_{AB(C)}^{ijk} Y_A^i (Y_B + Y_C)^j Y_C^k \qquad (26)$$

where the first term at the right-hand side is constant along the line "1-*O*" in Fig.1c and is equal to $\Delta g_{AB}$ in the binary system at point "1". The second term also represents the ternary interaction energy as before. One similar expression for $\Delta g_{CA}$ can be written as

$$\Delta g_{CA} = \left( \Delta g_{CA}^0 + \sum_{i+j\geq 1} q_{CA}^{ij} (Y_C + Y_B)^i Y_A^j \right) + \sum_{\substack{k\geq 1 \\ i\geq 0 \\ j\geq 0}} q_{CA(B)}^{ijk} (Y_C + Y_B)^i Y_A^j Y_B^k \qquad (27)$$

by substituting $\delta_{CA}^B = 1$ into Equation (15). Actually, the binary terms in Equations (26-27) are the same since the bond energies are assumed to be constant at constant $Y_A$. The ternary $\Delta g_{BC}$ expression is written in Equation (22) when the Kohler-type method is used for the interpolation.

Suppose that the binary $\Delta g_{AB}$ and $\Delta g_{CA}$ expressions are written in the polynomial of pair fractions. The ternary expressions can be given as

$$\Delta g_{AB} = \left( \Delta g_{AB}^0 + \sum_{i+j\geq 1} q_{AB}^{ij} X_{AA}^i (X_{BB} + X_{CC} + X_{BC})^j \right) + \sum_{\substack{k\geq 1 \\ i\geq 0 \\ j\geq 0}} q_{AB(C)}^{ijk} X_{AA}^i (X_{BB} + X_{CC} + X_{BC})^j Y_C^k \qquad (28)$$

$$\Delta g_{CA} = \left( \Delta g_{CA}^0 + \sum_{i+j\geq 1} q_{CA}^{ij} (X_{BB} + X_{CC} + X_{BC})^i X_{AA}^j \right) + \sum_{\substack{k\geq 1 \\ i\geq 0 \\ j\geq 0}} q_{CA(B)}^{ijk} (X_{BB} + X_{CC} + X_{BC})^i X_{AA}^j Y_B^k \qquad (29)$$

The ternary $\Delta g_{BC}$ expression is written in Equation (25) when the Kohler-type method is used for the interpolation. In the limit of ideality, Equations (28) and (29) reduce to Equations (26) and (27), respectively, while Equation (25) reduces to Equation (22), which becomes the well-known Kohler-Toop method for asymmetrical ternary systems.

### 3.4 The Muggianu method

To achieve the Muggianu interpolation, $\delta_{AB}^C = 1/2$ can be used in the generic method. Substituting it into Equation (14) gives

$$\Delta g_{AB} = \left(\Delta g_{AB}^0 + \Sigma_{(i+j)\geq 1} q_{AB}^{ij} (Y_A + \tfrac{1}{2}Y_C)^i (Y_B + \tfrac{1}{2}Y_C)^j\right) + \Sigma_{\substack{k\geq 1 \\ i\geq 0 \\ j\geq 0}} q_{AB(C)}^{ijk} (Y_A + \tfrac{1}{2}Y_C)^i (Y_B + \tfrac{1}{2}Y_C)^j Y_C^k \quad (30)$$

where the first term at the right-hand side is constant along the line "1-$O$" in Fig.1d and is equal to $\Delta g_{AB}$ in the binary system at point "1". This indicates that the binary (A-B) bond energy is constant when $Y_A - Y_B$ is constant. The second term is the ternary interaction. Substituting $\delta_{CA}^B = 1/2$ and $\delta_{BC}^A = 1/2$ into Equations (15-16) gives

$$\Delta g_{CA} = \left(\Delta g_{CA}^0 + \Sigma_{(i+j)\geq 1} q_{CA}^{ij} (Y_C + \tfrac{1}{2}Y_B)^i (Y_A + \tfrac{1}{2}Y_B)^j\right) + \Sigma_{\substack{k\geq 1 \\ i\geq 0 \\ j\geq 0}} q_{CA(B)}^{ijk} (Y_C + \tfrac{1}{2}Y_B)^i (Y_A + \tfrac{1}{2}Y_B)^j Y_B^k \quad (31)$$

$$\Delta g_{BC} = \left(\Delta g_{BC}^0 + \Sigma_{(i+j)\geq 1} q_{BC}^{ij} (Y_B + \tfrac{1}{2}Y_A)^i (Y_C + \tfrac{1}{2}Y_A)^j\right) + \Sigma_{\substack{k\geq 1 \\ i\geq 0 \\ j\geq 0}} q_{BC(A)}^{ijk} (Y_B + \tfrac{1}{2}Y_A)^i (Y_C + \tfrac{1}{2}Y_A)^j Y_A^k \quad (32)$$

where the first terms at the right-hand sides are constant along the lines "2-$O$" and "3-$O$" in Fig.1d and are equal to $\Delta g_{CA}$ and $\Delta g_{BC}$ in the binaries at point "2" and "3", respectively. This indicates the binary (C-A) and (B-C) bond energies are constant when $Y_C - Y_A$ and $Y_B - Y_C$ are constant. The second summations refer to the ternary interaction energies.

Suppose that the binary $\Delta g_{AB}$, $\Delta g_{CA}$ and $\Delta g_{BC}$ expressions are written in the polynomial of pair fractions. By taking account of those aforementioned facts, their ternary expressions can be given as

$$\Delta g_{AB} = \left(\Delta g_{AB}^0 + \Sigma_{(i+j)\geq 1} q_{AB}^{ij} \left(X_{AA} + \tfrac{1}{4}X_{CC} + \tfrac{1}{2}X_{AC}\right)^i \left(X_{BB} + \tfrac{1}{4}X_{CC} + \tfrac{1}{2}X_{BC}\right)^j\right) + \Sigma_{\substack{k\geq 1 \\ i\geq 0 \\ j\geq 0}} q_{AB(C)}^{ijk} \left(X_{AA} + \right.$$

$$\left. \tfrac{1}{4}X_{CC} + \tfrac{1}{2}X_{AC}\right)^i \left(X_{BB} + \tfrac{1}{4}X_{CC} + \tfrac{1}{2}X_{BC}\right)^j Y_C^k \quad (33)$$

$$\Delta g_{CA} = \left(\Delta g_{CA}^0 + \Sigma_{(i+j)\geq 1} q_{CA}^{ij} \left(X_{CC} + \tfrac{1}{4}X_{BB} + \tfrac{1}{2}X_{BC}\right)^i \left(X_{AA} + \tfrac{1}{4}X_{BB} + \tfrac{1}{2}X_{AB}\right)^j\right) + \Sigma_{\substack{k\geq 1 \\ i\geq 0 \\ j\geq 0}} q_{CA(B)}^{ijk} \left(X_{CC} + \right.$$

$$\left. \tfrac{1}{4}X_{BB} + \tfrac{1}{2}X_{BC}\right)^i \left(X_{AA} + \tfrac{1}{4}X_{BB} + \tfrac{1}{2}X_{AB}\right)^j Y_B^k \quad (34)$$

$$\Delta g_{BC} = \left(\Delta g_{BC}^0 + \Sigma_{(i+j)\geq 1} q_{BC}^{ij} \left(X_{BB} + \frac{1}{4}X_{AA} + \frac{1}{2}X_{AB}\right)^i \left(X_{CC} + \frac{1}{4}X_{AA} + \frac{1}{2}X_{CA}\right)^j\right) + \Sigma_{\substack{k\geq 1 \\ i\geq 0 \\ j\geq 0}} q_{BC(A)}^{ijk} \left(X_{BB} + \right.$$

$$\left.\frac{1}{4}X_{AA} + \frac{1}{2}X_{AB}\right)^i \left(X_{CC} + \frac{1}{4}X_{AA} + \frac{1}{2}X_{CA}\right)^j Y_A^k \tag{35}$$

In the limit of ideality, Equations (33-35) reduce to Equations (30-32), respectively, which becomes the well-known Muggianu method for symmetrical ternary systems.

It has been shown that [16], for systems with large composition-dependent deviations from ideality, the choice of a symmetric or an asymmetric model can often give very different results. An inappropriate choice may even give rise to artificial phase relations.

### 3.5 The integration method

When the interpolation parameters are assigned some special values, the generic method can be transformed into the reported geometric interpolation schemes. In addition, the generic method can also be applied to a ternary solution in case that all the reported geometric interpolation schemes are inappropriate to capture its thermodynamic features. However, once ternary experimental data are unavailable for a ternary solution, it is a great challenge to use the generic method to perform accurate predictions since there is no idea on how to determine the interpolation parameters. With the integration method [10], the challenge could be overcome. The interpolation parameters can be calculated by the so-called "deviation sum of squares", which are defined as

$$\eta_A = \int_0^1 (\Delta G_{AB} - \Delta G_{AC})^2 dn_A \tag{36}$$

$$\eta_B = \int_0^1 (\Delta G_{BA} - \Delta G_{BC})^2 dn_B \tag{37}$$

$$\eta_C = \int_0^1 (\Delta G_{CA} - \Delta G_{CB})^2 dn_C \tag{38}$$

where $\Delta G_{AB}$ (or $\Delta G_{BA}$), $\Delta G_{AC}$ (or $\Delta G_{CA}$) and $\Delta G_{BC}$ (or $\Delta G_{BC}$) represent the Gibbs energy of mixing in the binary A-B, A-C and B-C solutions, respectively. $n_A$, $n_B$ and $n_C$ refer respectively to the number of moles of component A in the binary A-B and A-C solutions, that of component B in the binary B-C and B-A solutions, and that of component C in the binary C-A and C-B solutions.

Clearly, if component B is similar to component C thermodynamically, the value of $\eta_A$ will approach zero, otherwise it is expected to be a positive nonzero. Similar scenarios are found for $\eta_B$ and $\eta_C$. The interpolation parameters can be obtained as

$$\delta_{AB}^C = \frac{\eta_A}{\eta_A+\eta_B} = 1 - \delta_{BA}^C \qquad (39)$$

$$\delta_{BC}^A = \frac{\eta_B}{\eta_B+\eta_C} = 1 - \delta_{CB}^A \qquad (40)$$

$$\delta_{CA}^B = \frac{\eta_C}{\eta_C+\eta_A} = 1 - \delta_{AC}^B \qquad (41)$$

where $\delta$ is also termed as "similarity coefficient" [10]. This indicates that the identity between components C and B causes $\eta_A = 0$ and thus $\delta_{AB}^C = 0$ while the identity between components C and A leads to $\eta_B = 0$ and thus $\delta_{AB}^C = 1$. Therefore, from the $\delta_{AB}^C$ value, one can judge which of components A and B is more similar to component C. Similar situations can be found for $\delta_{BC}^A$ and $\delta_{CA}^B$.

In order to obtain the interpolation parameters, the $\eta$ values have to be firstly calculated using Equations (36-38). Since the Gibbs energies of mixing are now expressed in the quasichemical formalism, it is infeasible to analytically solve the integrals due to the unsolved internal variables (pair numbers). In sub-binary solutions, those equilibrium pair numbers are initially obtained by minimizing the Gibbs energies at constant compositions, temperatures and pressures, and they are then taken back into the Gibbs energy expressions. However, it is not sure if the equilibrium pair numbers could be taken into Equations (36-38) to obtain the optimal $\eta$ values. This uncertainty could be well clarified as follows. Let $F(n_A, n_{AB}, n_{AC})$ be the integrand as

$$F(n_A, n_{AB}, n_{AC}) = [\Delta G_{AB}(n_A, n_{AB}) - \Delta G_{13}(n_A, n_{AC})]^2 \qquad (42)$$

$\eta_A$ is then expressed as

$$\eta_A = \int_0^1 F(n_A, n_{AB}, n_{AC}) dn_A \qquad (43)$$

Minimization of $\eta_A$ yields

$$\frac{\partial \eta_A}{\partial n_{AB}} = \int_{0=f(n_{AB},n_{AC})}^{1=g(n_{AB},n_{AC})} \frac{\partial F(n_A,n_{AB},n_{AC})}{\partial n_{AB}} dn_A + \left[\frac{\partial g}{\partial n_{AB}} - \frac{\partial f}{\partial n_{AB}}\right] F(n_A, n_{AB}, n_{AC}) = 0 \qquad (44)$$

$$\frac{\partial \eta_A}{\partial n_{AC}} = \int_{0=f(n_{AB},n_{AC})}^{1=g(n_{AB},n_{AC})} \frac{\partial F(n_A,n_{AB},n_{AC})}{\partial n_{AC}} dn_A + \left[\frac{\partial g}{\partial n_{AC}} - \frac{\partial f}{\partial n_{AC}}\right] F(n_A, n_{AB}, n_{AC}) = 0 \qquad (45)$$

where at the right-hand side the second terms are always zero. This indicates that the following equations must hold,

$$\frac{\partial F(n_A,n_{AB},n_{AC})}{\partial n_{AB}} = 2[\Delta G_{AB}(n_A, n_{AB}) - \Delta G_{AC}(n_A, n_{AC})]\left[\frac{\partial \Delta G_{AB}(n_{AB},n_A)}{\partial n_{AB}}\right] = 0 \qquad (46)$$

$$\frac{\partial F(n_A,n_{AB},n_{AC})}{\partial n_{AC}} = 2[\Delta G_{AC}(n_A, n_{AC}) - \Delta G_{AB}(n_A, n_{AB})]\left[\frac{\partial \Delta G_{AC}(n_{AC},n_A)}{\partial n_{AC}}\right] = 0 \qquad (47)$$

because $n_A$ is an arbitrary value ranging from 0 to 1 in Equations (46-47). In general, the deviation between $\Delta G_{AB}$ and $\Delta G_{AC}$ should not always disappear, which results in the following expressions,

$$\frac{\partial \Delta G_{AB}(n_{AB}, n_A)}{\partial n_{AB}} = 0 \quad (48)$$

$$\frac{\partial \Delta G_{AC}(n_{AC}, n_A)}{\partial n_{AC}} = 0 \quad (49)$$

It is evident that Equations (48-49) are the right minimization procedures to obtain the equilibrium pair numbers for the binary A-B and A-C solutions. This indicates that the $\eta$ values could be calculated by just taking the equilibrium pair numbers obtained from the sub-binary solutions into the integral equations. The integral equation can be approximated by its closely related sum,

$$\eta_A = \int_0^1 (\Delta G_{AB} - \Delta G_{AC})^2 dn_A \cong \sum_{i=1}^{1/\Delta h - 1} \Delta h [\Delta G_{AB}(i\Delta h, n_{AB}^*) - \Delta G_{AC}(i\Delta h, n_{AC}^*)]^2 \quad (50)$$

where $n_{AB}^*$ and $n_{AC}^*$ are the equilibrium numbers of moles of the A-B and A-C pairs in the sub-binary A-B and A-C solutions at the composition $i\Delta h$, respectively. It is found that $\Delta h$ equal to 0.01 could yield good approximations. The similar approximations could be used for calculating $\eta_B$ and $\eta_C$. All the interpolation parameters are then determined by the $\eta$ values. It is obvious that $\delta$ can be a function of temperature since $\eta$ may be changeable against temperature. Similar to other thermodynamic parameters, such as Gibbs energies of pure components, interaction parameters between components, magnetic parameters, *etc.*, the interpolation parameters can be placed in thermodynamic database so that the current algorithm widely used in thermodynamic software is convenient to compile the present thermodynamic model.

## 4 Case study

In this section, the Li-Cu, Li-Sn and Cu-Sn liquids were first thermodynamically optimized using their binary experimental data. All optimizations and calculations were conducted by our own homemade code. The optimized model parameters were then substituted into the GBEF within the MQMPA to give thermodynamic descriptions of the Cu-Li-Sn liquid. By varying the interpolation parameters ($\delta_{LiCu}^{Sn}$, $\delta_{SnLi}^{Cu}$ and $\delta_{CuSn}^{Li}$), the present model provided the Cu-Li-Sn liquid various thermodynamic predictions. The reliability of these predictions was then evaluated by the ternary data in experimental thermochemistry and phase equilibria.

Table 1 Coordination numbers and model parameters used in the Li-Sn, Cu-Sn and Li-Cu systems (T in Kelvin, $\Delta g$ in J/mol)

| Systems / parameters | Sn-Li | Li-Cu | Cu-Sn |
|---|---|---|---|
| Coordination numbers | $Z^{Li}_{SnLi} = 1.5$  $Z^{Sn}_{SnLi} = 6$ | $Z^{Li}_{LiCu} = 6$  $Z^{Cu}_{LiCu} = 6$ | $Z^{Cu}_{CuSn} = 6$  $Z^{Sn}_{CuSn} = 6$ |
| | $Z^{Li}_{LiLi} = 6$ | $Z^{Cu}_{CuCu} = 6$ | $Z^{Sn}_{SnSn} = 6$ |
| Model parameters | $\Delta g^0_{SnLi} = -65000$ | $\Delta g^0_{LiCu} = 9834.27 - 8T$  $q^{10}_{LiCu} = -2713.6 + 1.0T$ | $\Delta g^0_{CuSn} = 356.9 - 0.8504T$ |
| | $q^{10}_{SnLi} = -17000$ | | $q^{10}_{CuSn} = -14527.57 + 0.1230T$ |
| | $q^{01}_{SnLi} = 0$ | $q^{01}_{LiCu} = -16703.2$ | $q^{01}_{CuSn} = -146.77 - 3.3516T$ |

According to Equation (4), the binary bond energy expressions are given as

$$\Delta g_{SnLi} = \Delta g^0_{SnLi} + q^{10}_{SnLi}X_{SnSn} + q^{01}_{SnLi}X_{LiLi} \tag{51}$$

$$\Delta g_{LiCu} = \Delta g^0_{LiCu} + q^{10}_{LiCu}X_{LiLi} + q^{01}_{LiCu}X_{CuCu} \tag{52}$$

$$\Delta g_{CuSn} = \Delta g^0_{CuSn} + q^{10}_{CuSn}X_{CuCu} + q^{01}_{CuSn}X_{SnSn} \tag{53}$$

where the model parameters and the defined coordination numbers are listed in Table 1. Fig.2 shows the mixing enthalpy of the Sn-Li liquid calculated by the MQMPA and ASM along with the experimental data [17]. All the calculated curves show deep minima near the SnLi$_4$ position, indicating that the Li and Sn atoms form a strong SRO in the molten alloy. The ratio $Z^{Sn}_{SnLi}/Z^{Li}_{SnLi} = 4$ is chosen to define the SRO composition. Results of the calculations for the Li-Cu and Cu-Sn liquids along with the experimental data [18-19] are presented in Figs.3-4 where both of the two curves display S-shaped profiles. The two liquids are close to regular solutions wherein all atoms tend to be randomly distributed. The same coordination numbers are thus defined for all atoms in the Li-Cu and Cu-Sn liquids.

According to Equations (17-19), the model parameters used in Equations (51-53) can be directly used to derive the ternary bond energy expressions,

$$\Delta g_{LiCu} = \Delta g^0_{LiCu} + q^{10}_{LiCu}\left(X_{LiLi} + \delta^{Sn}_{LiCu}{}^2 X_{SnSn} + \delta^{Sn}_{LiCu}X_{LiSn}\right) + q^{01}_{LiCu}\left(X_{CuCu} + \delta^{Sn}_{CuLi}{}^2 X_{SnSn} + \delta^{Sn}_{CuLi}X_{CuSn}\right) \tag{54}$$

$$\Delta g_{CuSn} = \Delta g^0_{CuSn} + q^{10}_{CuSn}\left(X_{CuCu} + \delta^{Li}_{CuSn}{}^2 X_{LiLi} + \delta^{Li}_{CuSn}X_{CuLi}\right) + q^{01}_{CuSn}\left(X_{SnSn} + \delta^{Li}_{SnCu}{}^2 X_{LiLi} + \delta^{Li}_{SnCu}X_{SnLi}\right) \tag{55}$$

$$\Delta g_{SnLi} = \Delta g^0_{SnLi} + q^{10}_{SnLi}\left(X_{SnSn} + \delta^{Cu}_{SnLi}{}^2 X_{CuCu} + \delta^{Cu}_{SnLi}X_{CuSn}\right) \tag{56}$$

where the interpolation parameters are not confined to invariable values in a specific geometric interpolation method. The optimal parameters can be found by the best fit to all measured enthalpy of mixing data in the Cu-Li-Sn liquid. It is noticeable that no ternary interactions are considered in the present modeling. Substituting Equations (54-56) into Equation (1), one can calculate the phase equilibria and thermodynamic properties of the Cu-Li-Sn liquid.

By means of drop calorimetry, Fürtauer [17] has carefully measured the mixing enthalpy of the Cu-Li-Sn liquid along five isopleth sections. These experimental data are very useful to obtain the optimal interpolation parameters and meanwhile to examine the calculated results from other geometric interpolation methods. Fig.5 displays the mixing enthalpy in the ternary liquid calculated and measured along the five isopleth sections. By employing $\delta_{LiCu}^{Sn} = Y_{Li}/(Y_{Li} + Y_{Cu})$, $\delta_{SnLi}^{Cu} = Y_{Sn}/(Y_{Li} + Y_{Sn})$ and $\delta_{CuSn}^{Li} = Y_{Cu}/(Y_{Cu} + Y_{Sn})$, the symmetrical Kohler method is realized to predict the thermodynamic properties of the ternary solution. By assigning $\delta_{LiCu}^{Sn} = 1.0$, $\delta_{SnLi}^{Cu} = Y_{Sn}/(Y_{Li} + Y_{Sn})$ and $\delta_{CuSn}^{Li} = 0.0$, the asymmetrical Kohler-Toop method is achieved and Cu is chosen as the asymmetrical component. By using $\delta_{LiCu}^{Sn} = \delta_{SnLi}^{Cu} = \delta_{CuSn}^{Li} = 0.5$, the symmetrical Muggianu method is implemented. With $\delta_{LiCu}^{Sn} = 0.9989$, $\delta_{SnLi}^{Cu} = 0.4635$ and $\delta_{CuSn}^{Li} = 0.0012$, the integration method is carried out through Equation (50). With the best fit to the experimental mixing enthalpy [17], $\delta_{LiCu}^{Sn} = 1.0$, $\delta_{SnLi}^{Cu} = 1.0$ and $\delta_{CuSn}^{Li} = 0.0$ are found and used in the GBEF. It is observed from Fig.5 (a), Fig.5 (b) and Fig.5 (c) that the exhibited curves are similar to those for the Li-Sn liquid on which the deep minima appear. This happened since Li was experimentally dropped into the Cu-Sn liquid. It is seen from Figs.5 (d) and (e) that the displayed curves show S-shaped profile and are similar to those for the Cu-Li and Cu-Sn liquids. This occurred because Cu was experimentally dropped into the Li-Sn liquid.

Similar analyses were made by Li et al. [20] who introduced the complex Li$_4$Sn species in the ASM to treat the Li-Sn liquid with strong SRO, and used the substitutional solution model with the Redlich-Kister polynomials to describe the Cu-Sn and Li-Cu liquids. Li et al. [20] then calculated the thermodynamic properties of the Cu-Li-Sn liquid by employing the symmetric Muggianu method, the asymmetric Muggianu-Toop method and an interpolation method based on the calculation of the global Gibbs free energy minimum for the ternary system. Their model also ignored ternary interactions. However, a good reproduction of the ternary experimental data could only be achieved by the last interpolation method. We have reproduced their calculation results by using their liquid model (Li, Cu, Sn and Li$_4$Sn) with the last interpolation method. It is obvious from Figs.5a-5c that the calculation method of Li et al [20] results in much sharper curves. Meanwhile, their method is unable to predict a directional liquid-liquid miscibility gap along the Cu-Li$_4$Sn line, which can generally give a two-liquid region at the center zone of the Gibbs triangle. This is not abnormal [21] since Cu and Li$_4$Sn are randomly mixed in their liquid model. The directional two-liquid region can be reasonably predicted along the Cu-Li$_4$Sn line using the present model without ternary interactions. Fig.6 clearly shows the comparison between the model of Li et al. [20] and the present model. In the ternary liquid, atom Li always tends to stay with atom Sn, leaving atom Cu lone. This renders the Cu-Cu pair and Li-Sn pair to be dominant in the Cu-Li-Sn liquid, resulting in the formation of the two-liquid region along the Cu-Li$_4$Sn direction. The present model can effectively capture this phenomenon since it has inherited the essence of the MQMPA.

Fürtauer et al. have reported four isothermal sections [22] and nine isopleth sections plus one liquidus project [23] in the Cu-Li-Sn system according to their experiments. The isothermal sections show that there only exist stable terminal single-liquid zones between 300-600°C. A stable liquid-liquid miscibility gap is schematically portrayed nearly along the Cu-Li$_4$Sn line on the liquidus project. After carefully analyzing experimental results, Fürtauer et al. confirmed that the critical liquidus temperature of the two-liquid zone is about 850 °C, above which the system enters the single-liquid region. However, as Fürtauer et al. [23] pointed out, the two-liquid zone could not be established in detail based on their experimental data. They used question marks in the isopleth sections to indicate where the liquid-liquid miscibility gap possibly appears. From the isopleth sections, it is shown that the two-liquid zone appears to be stable above 800 °C. Below this temperature, the two-liquid zone disappears while solid phases appear to be stable. By using the GBEF within the MQMPA, it is predicted that the two-liquid field can be stable above 800 °C and will not disappear until 1100 °C by utilizing all the special interpolation parameters (referring to the Kohler, Toop, Muggianu and the integration methods) but those optimized by the measured enthalpy of mixing data. This means the predictions from the former interpolations enlarge the two-liquid zone because the solution thermodynamics are characterized with positive deviations from ideality; the prediction from the latter interpolation is in the opposite situation. This is clearly reflected from Fig.6. In order to drive thermodynamic calculations to be consistent with the phase-equilibria experiments [23], ternary interactions must be included in the GBEF to adjust the stability range of the two-liquid zone. It has been found that the ternary interaction coefficient $q^{011}_{LiSn(Cu)}$ is more effective to affect the stability of the two-liquid region. If all calculations are expected to have the two-liquid zone within the stability range from 800 to 850 °C, the coefficient should be assigned -5200, -5000, -3000, -1800 and 1800 J/mol for the Kohler, Kohler-Toop, Muggianu, integration and optimization methods, respectively. It is obvious that the integration and optimization methods can employ weaker ternary interactions to achieve reliable thermodynamic calculations for the Cu-Li-Sn liquid.

The GBEF enables the MQMPA to be applicable to ternary solutions with various configurations. It can be conveniently transformed into the formulae derived from the Kohler, Muggianu and Toop methods using the special interpolation parameters. In fact, the interpolation parameters should be system dependent. Hence, ternary experimental data is necessary to determine the interpolation parameters. Normally, negligible ternary interactions are required to provide fine reproduction of all ternary data in thermochemistry and phase equilibria. Once ternary experimental data are unavailable, the integration method [10] can be employed to reasonably obtain the interpolation parameters. Moreover, the GBEF is more concise than the formulae derived by Pelton et al. [9] from a combinatorial Kohler-Toop method. Meanwhile, the GBEF can provide more interpolation methods to characterize ternary solution thermodynamics.

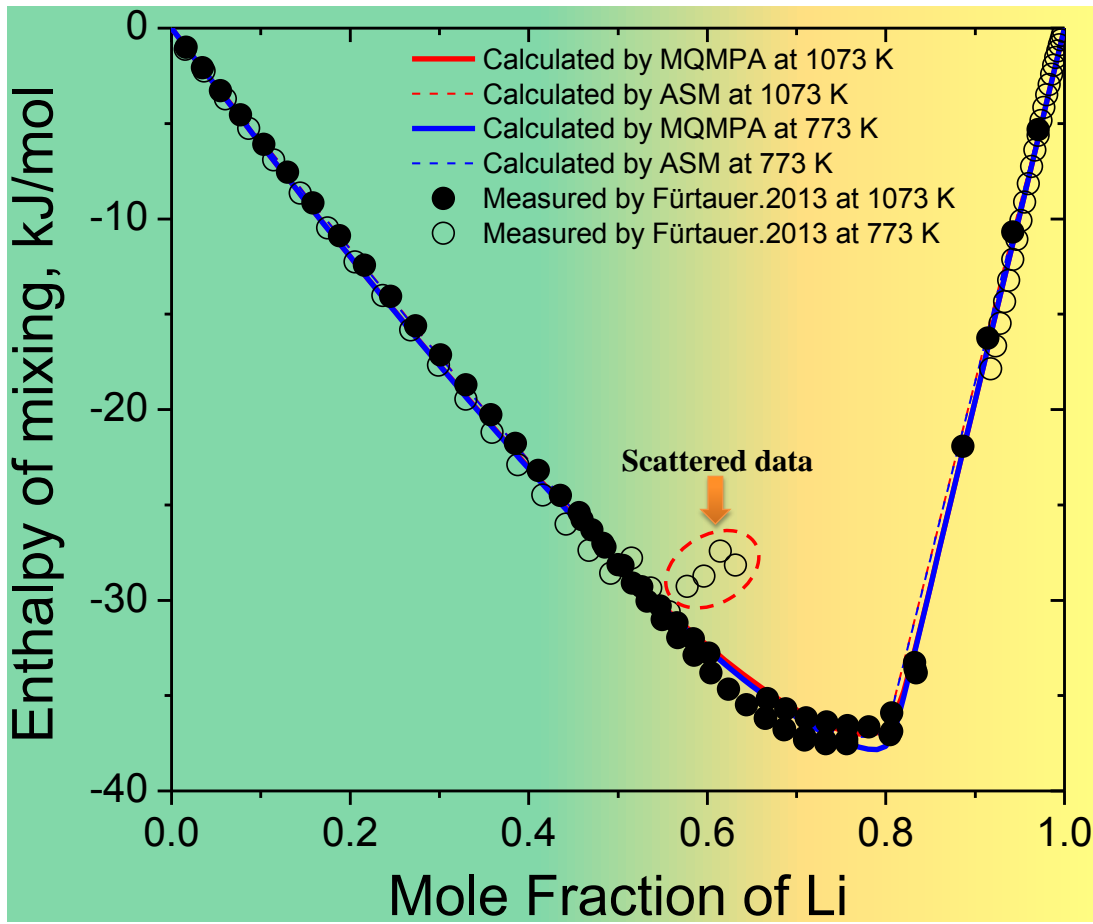

Fig. 2. Calculated mixing enthalpy for the Li-Sn liquid along with the experimental data [17]

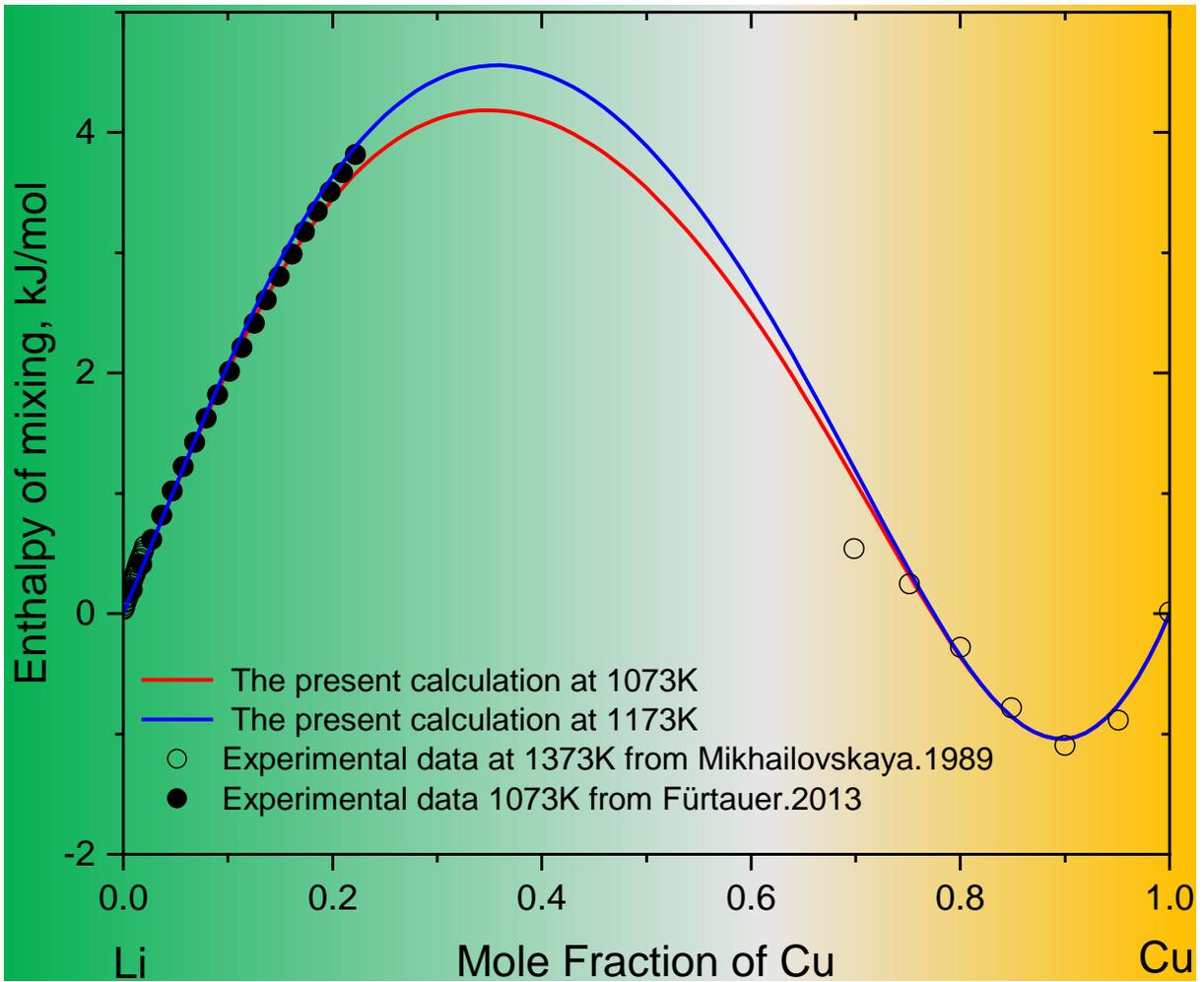

Fig. 3. Calculated mixing enthalpy for the Li-Cu liquid along with the experimental data [17-18]

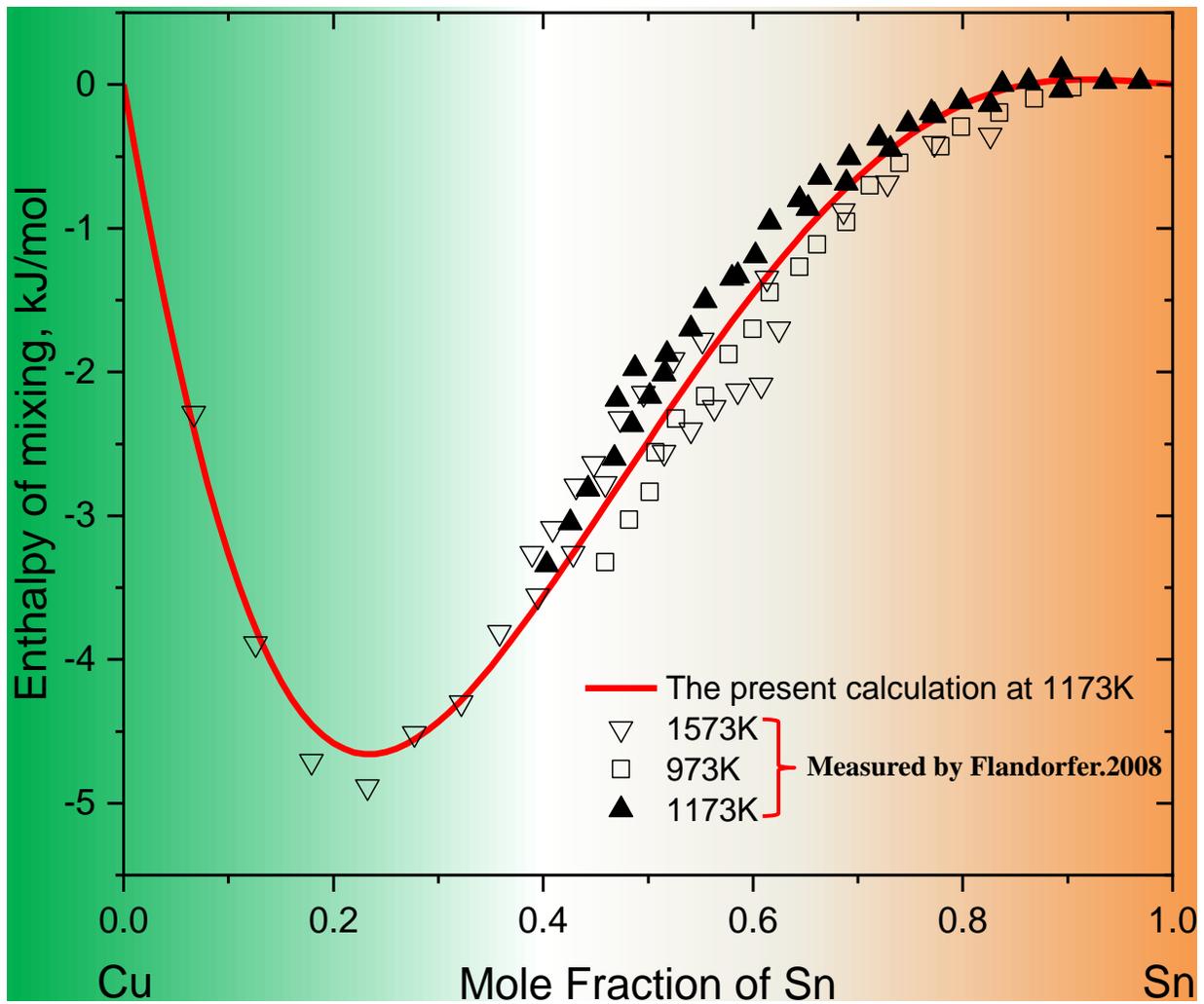

Fig. 4. Calculated mixing enthalpy for the Cu-Sn liquid along with the experimental data [19]

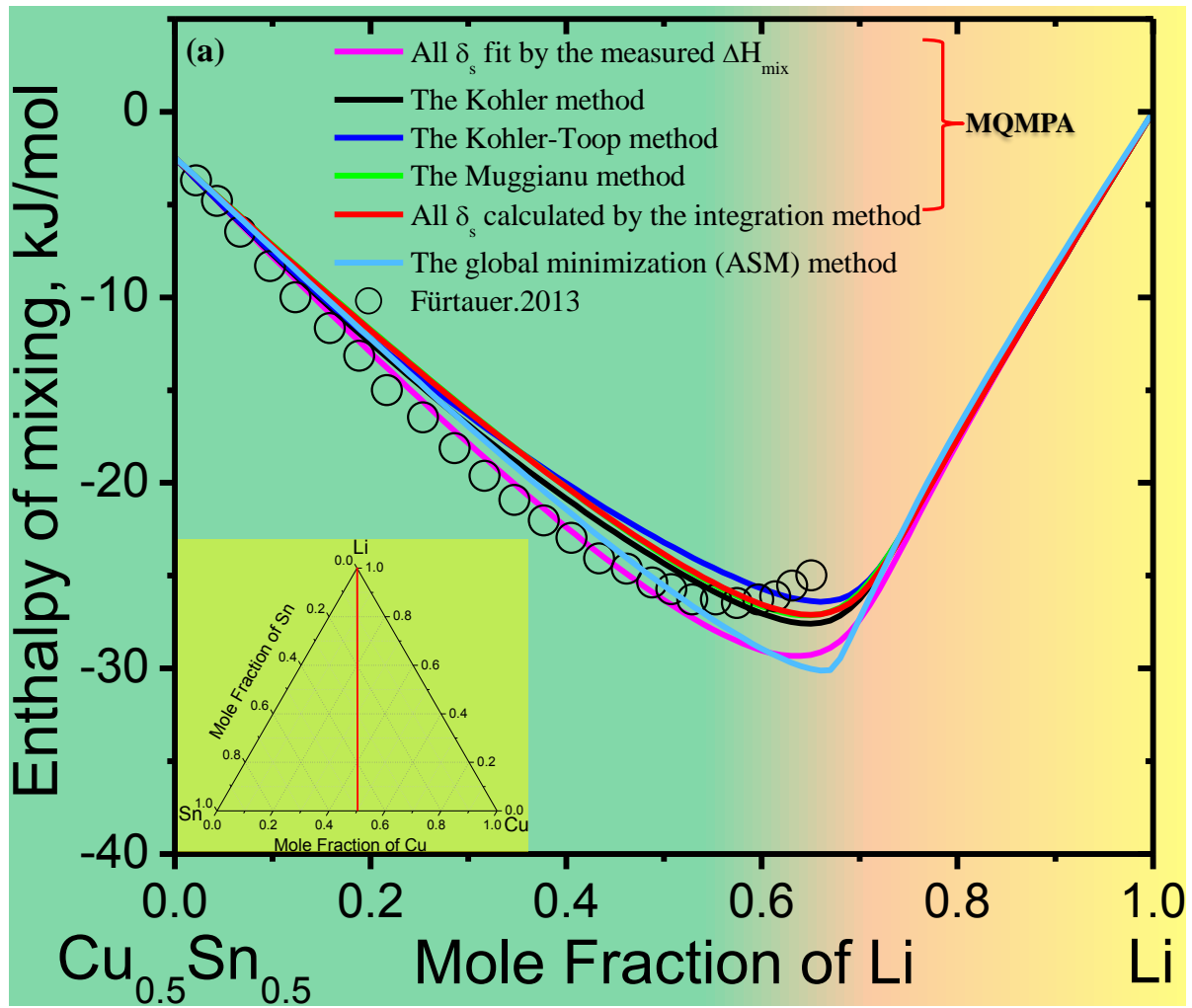

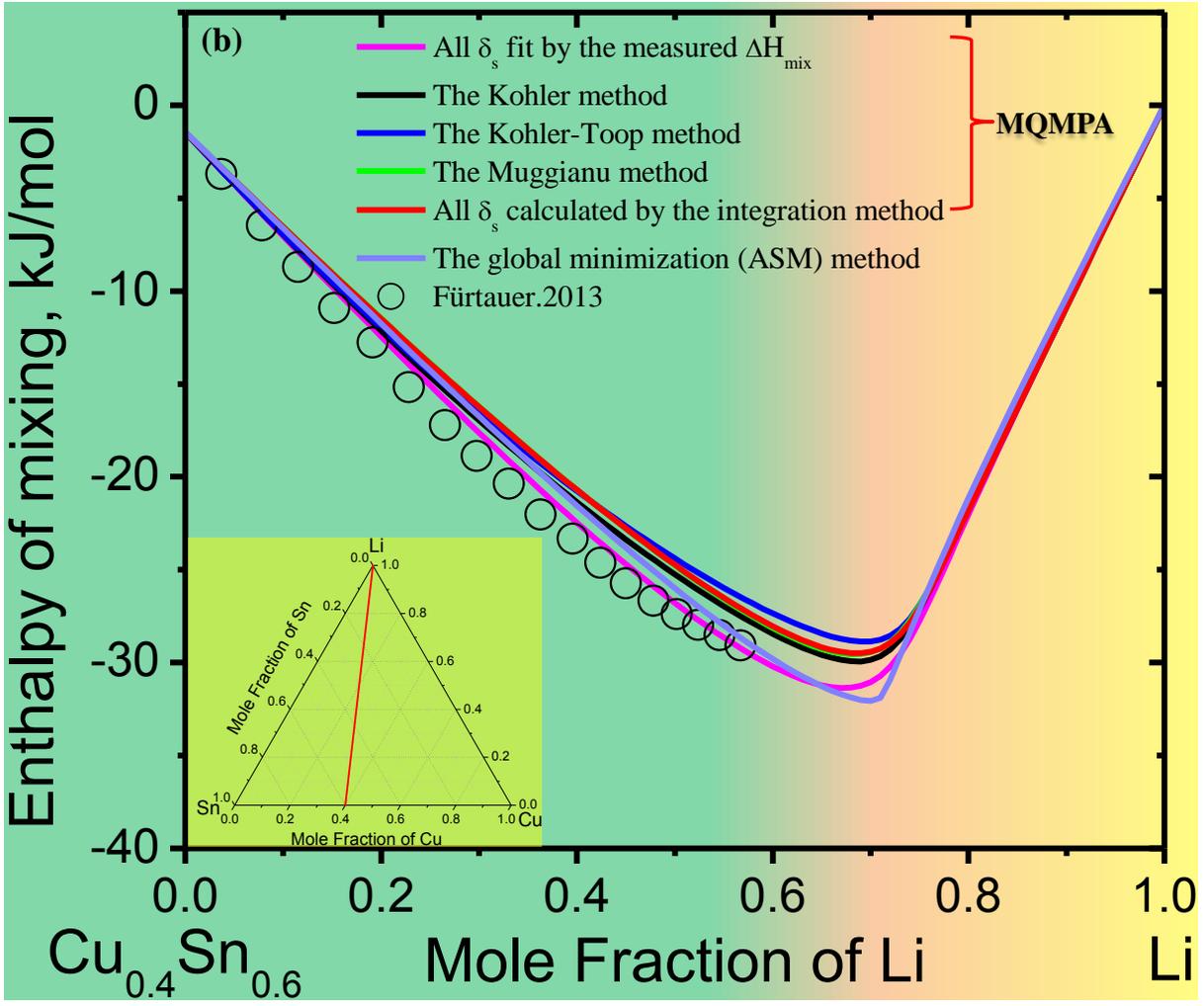

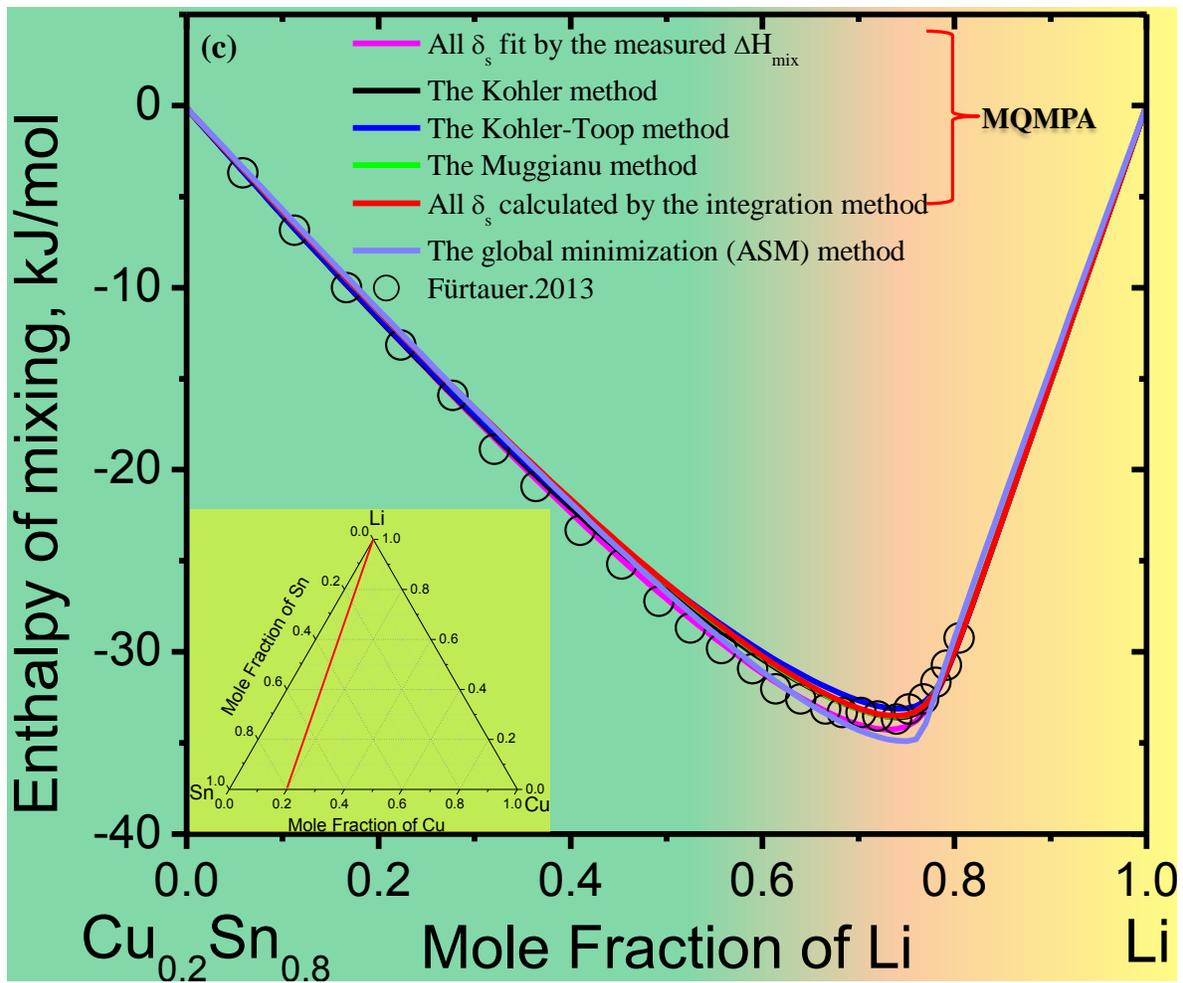

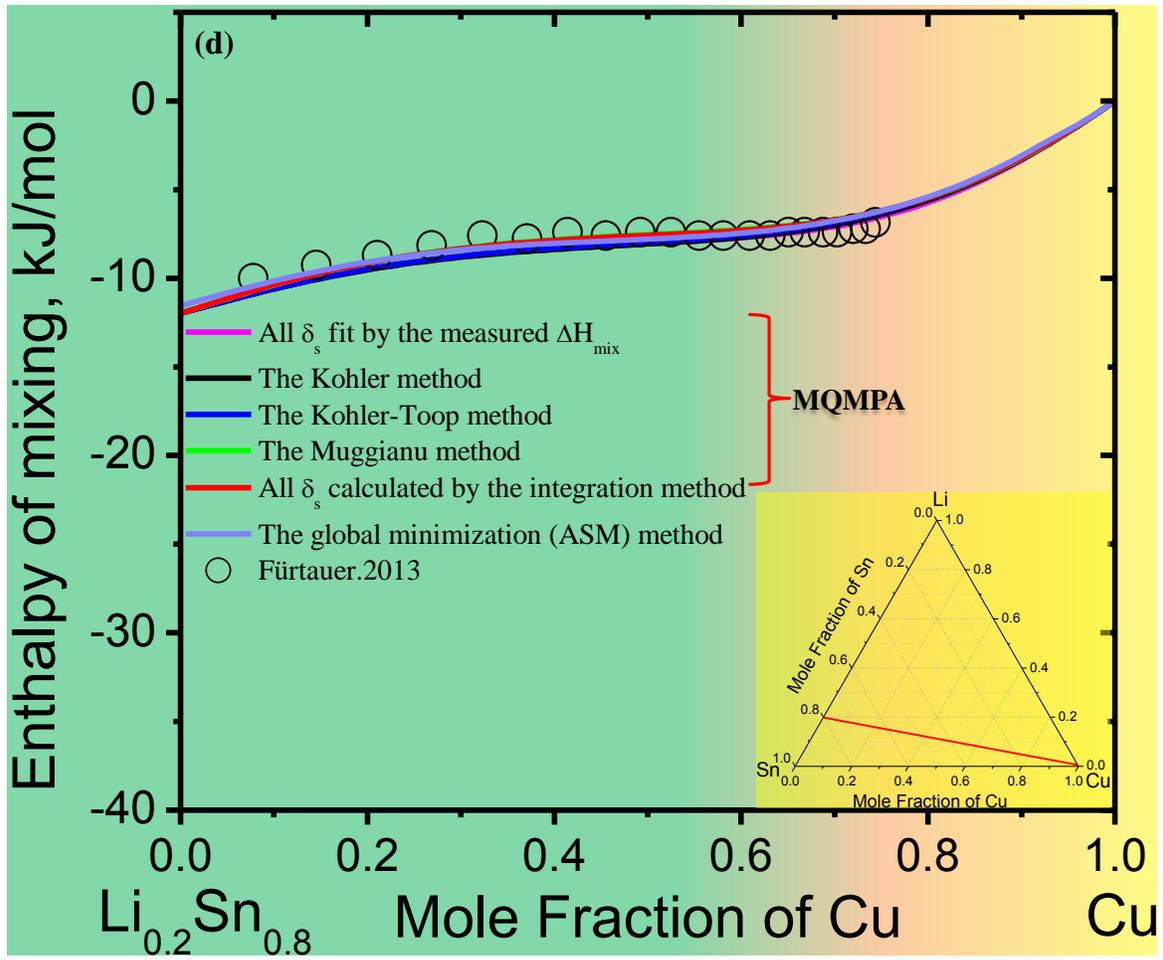

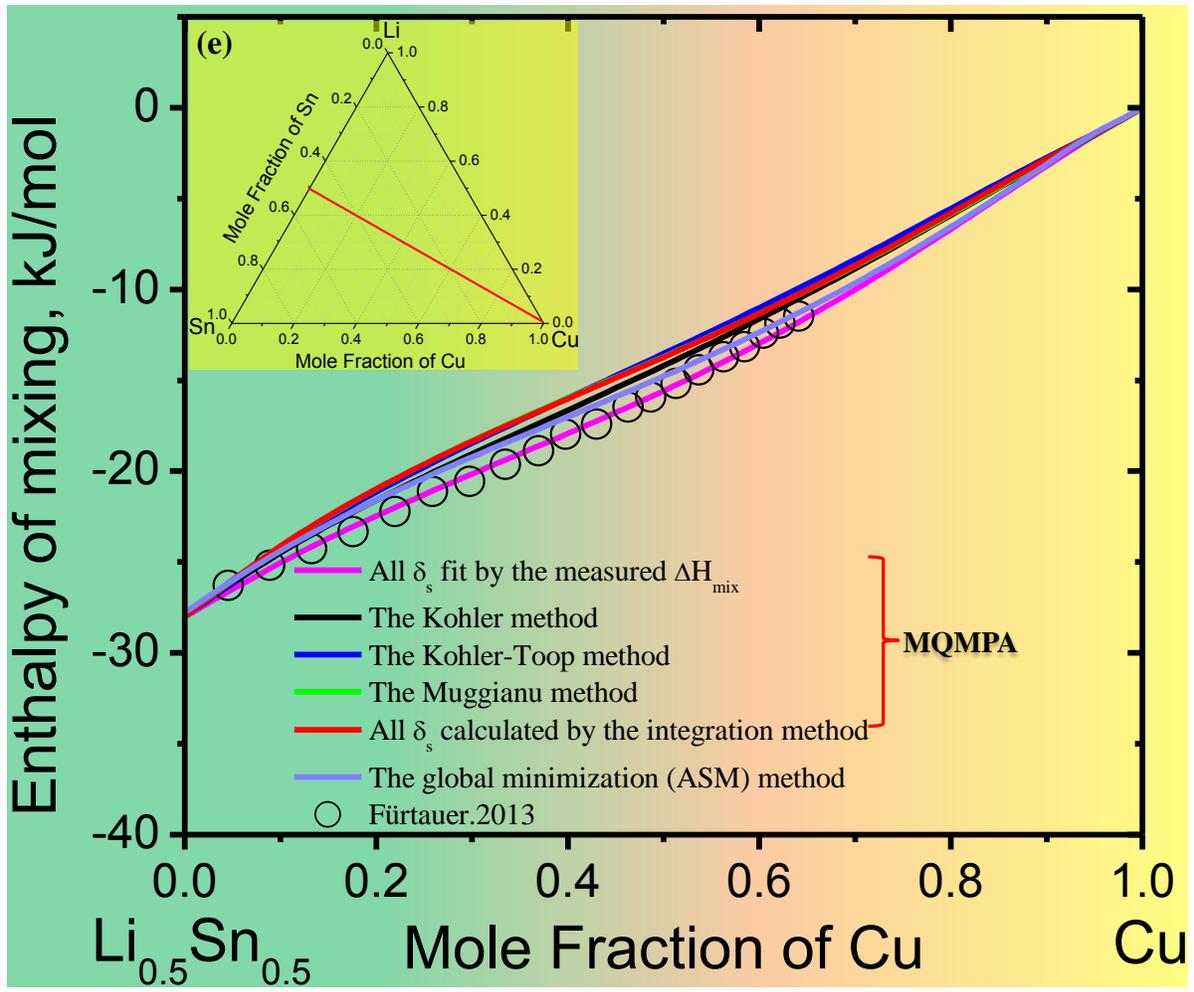

Fig. 5. Calculated mixing enthalpy in the Cu-Li-Sn liquid at 1073K using various models without ternary interactions

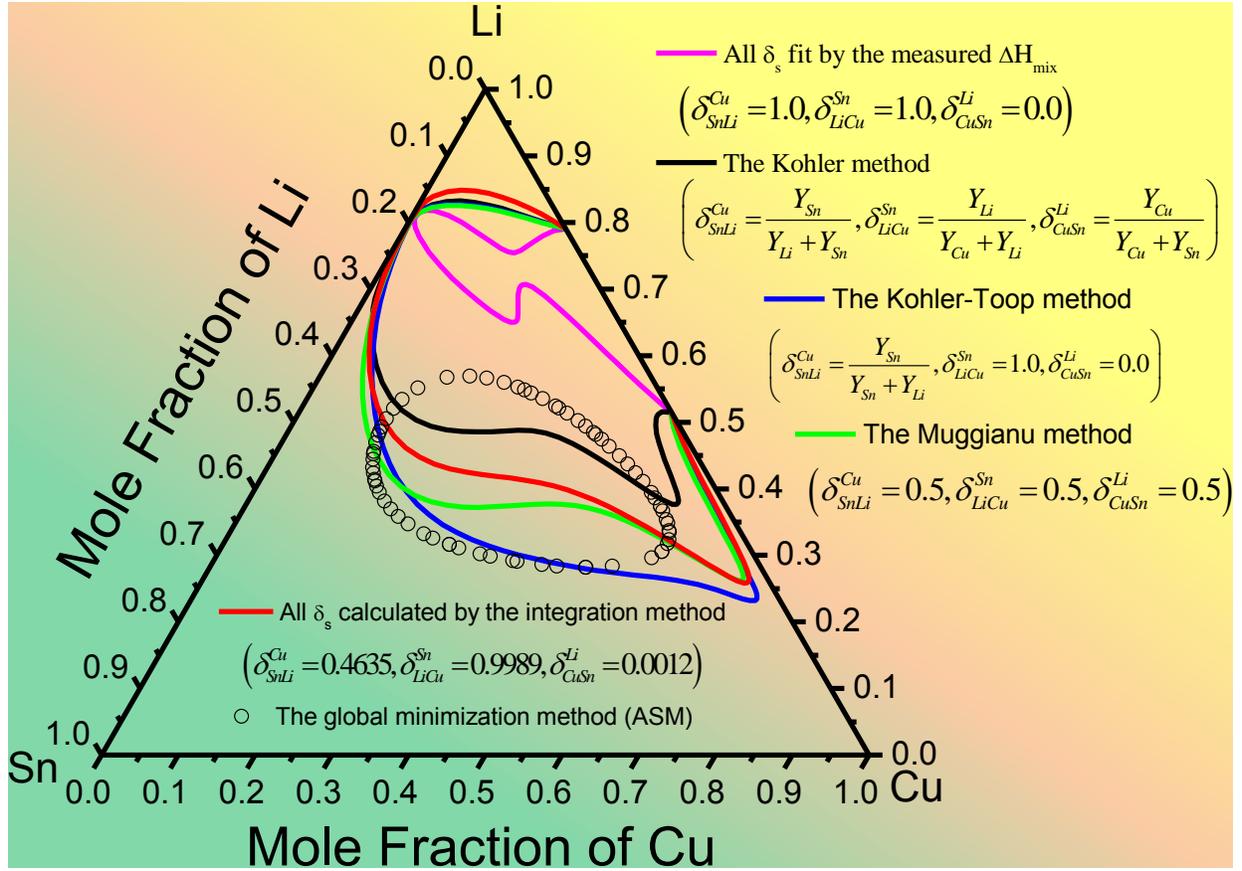

Fig. 6. Metastable liquid-liquid miscibility gap predicted at 773 K using various models without ternary interactions

## 5 Concluding remarks

A binary solution with SRO manifests characteristic solution thermodynamics. The thermodynamic features cannot be effectively represented by the solution models in the Bragg-Williams approximation but can be reasonably captured by the MQMPA. Once the MQMPA is used to treat ternary solutions, a proper geometric interpolation method must be required to transform the bond energy expression from binary to ternary formalism. Such transformation has been achieved in the present work using a generic geometric interpolation approach, leading to the successful development of the GBEF within the MQMPA. The GBEF can reduce to the formulae built by the widely used Kohler, Toop and Muggianu methods with some special interpolation parameters. The interpolation parameters can be calculated by the integration method as well as be optimized by ternary experimental data. These adjustable interpolation parameters provide the GBEF great flexibility to treat the thermodynamics of a ternary solution with complex configurations. Moreover, the GBEF is much more concise than that built by a combinatorial Kohler-Toop approach, facilitating the code implementation of this model.

The Cu-Li-Sn liquid with complex configurations is utilized to validate the reliability and evaluate the effectiveness of the present model. The predictive range can be clearly shown by just changing the interpolation parameters and driving the interpolation scheme to freely shift among all reported geometric solution models. The optimal interpolation parameters were eventually identified, which enable the present model to better reproduce the mixing enthalpy of the ternary liquid. The present model can also reasonably predict a potential liquid-liquid miscibility gap along the Cu-Li$_4$Sn direction, which is hardly described by the ASM without ternary interactions.

## Acknowledgements

This work is supported by the Program for Professor of Special Appointment (Eastern Scholar) at Shanghai Institutions of Higher Learning (TP2020034, TP2019041), National Natural Science Foundation of China (Grant Nos. 52022054 and 51974181).